KHz
\documentclass[%
 reprint,
 amsmath,amssymb,https://www.overleaf.com/5798916374sfbsjrjcnjgm
 aps,
]{revtex4-1}

\usepackage{graphicx}
\usepackage{dcolumn}
\usepackage{bm}
\usepackage{multirow}

\begin{document}

\preprint{APS/123-QED}

\title{Investigating students seriousness during selected conceptual inventory surveys}

\author{David P. Waters}
  \affiliation{Department of Basic Sciences, St. Louis College of Pharmacy, St. Louis, Missouri 63110, USA.}
\author{Dragos Amarie}
  \affiliation{Department of Physics and Astronomy, Georgia Southern University, Statesboro, Georgia 30460, USA.}
\author{Rebecca A. Booth}
 \affiliation{Calgary Board of Education, Calgary, Alberta, Canada.}
\author{Christopher Conover}
 \affiliation{Department of Physics and Astronomy, Kansas State University, Manhattan, Kansas 66506, USA.}
 \author{Eleanor C. Sayre}
 \affiliation{Department of Physics and Astronomy, Kansas State University, Manhattan, Kansas 66506, USA.}

\date{\today}

\begin{abstract}

Conceptual inventory surveys are routinely used in education research to identify student learning needs and assess instructional practices. Students might not fully engage with these instruments because of the low stakes attached to them. This paper explores tests that can be used to estimate the percentage  of students in a population who might not have taken such surveys seriously. These three seriousness tests are the pattern recognition test, the easy questions test, and the uncommon answers test. These three tests are applied to sets of students who were assessed either by the Force Concept Inventory, the Conceptual Survey of Electricity and Magnetism, or the Brief Electricity and Magnetism Assessment. The results of our investigation are compared to computer simulated populations of random answers.

\end{abstract}

\maketitle


\section{\label{sec:level1}Introduction}

Conceptual Inventories (CIs) came out of our necessity to quantify students' understanding of concepts and their progress in class by monitoring learning gains. \cite{r11}. The physics education research that followed has driven modern teaching with a focus on developing novel methods to stimulate students' understanding, and has also redefined our learning goals \cite{r21}. Halloun and Hestenes raised the concern that traditional instruction marginally affects students' understanding while their common sense beliefs usually contradict the laws of physics \cite{r31,r41}. Their Force Concept Inventory (FCI) survey arrives as a first tool to measure students’ mastery of force concepts widely taught in the first semester of physics \cite{r51}. Since then, CIs have gained widespread use in physics and astronomy education \cite{r52, r61, r62}, as well as many other disciplines of STEM \cite{r71,r81,r91,r101,r111,r121,r131,r141,r151,r161,r171,r181,r191,r201}.

Since CIs became more useful to instructors, they started to be used as research-based assessment instruments (RBAIs) in education research \cite{r211}. RBAIs are multiple-choice but carefully designed survey to provide insight into students' attitudes and understanding. Over time, RBAIs have undergone different rounds of scrutiny and validation \cite{r221}. When RBAI data is collected regularly, they could be valuable measuring tools by providing standardized comparisons among institutions, instructors, teaching methods, and over multiple implementations of the same course. They also allow us to track trends and investigate correlations over time \cite{r231,r241}. The physics education research that has followed from the use of RBAIs has driven physics instructors toward developing and implementing novel methods for increasing students’ understanding as well as toward redefining student learning goals  \cite{r21}. PhysPort, an online resource for instructors interested in implementing research-based physics teaching practices in their classrooms, currently provides 92 RBAIs with diverse foci, including content knowledge, problem-solving, scientific reasoning, lab skills, beliefs and attitudes, and interactive teaching \cite{r251}. 

Among the RBAIs available on PhysPort are the Force Concept Inventory (FCI), the Brief Electricity and Magnetism Assessment (BEMA), and the Conceptual Survey of Electricity and Magnetism (CSEM). The FCI is a 30-question RBAI used to measure students’ mastery of the mechanics concepts widely taught in a first-semester introductory physics course \cite{r51}. The FCI is among the most popular RBAIs, with extensive research on its efficacy as effects on instruction as well as many translations into different languages and formats. In particular, the FCI has been investigated by Hestenes et al. who interviewed students and instructors to confirm that surveyed individuals correctly understood the wording and the pictographs \cite{r51,r261}, whereas Stewart \textit{et al.} confirms that test scores are not particularly context dependent \cite{r271}. Version H of the CSEM, published by Maloney, O'Kuma, Hieggelke, and Van Heuvelen in 2001 \cite{r281}, is a 32-question RBAI used to measure student conceptual understanding of electricity and magnetism at an introductory undergraduate level. The BEMA is a 31-question RBAI also designed to assess conceptual understanding of electromagnetism.

\subsection*{Main concerns with RBAI}

From the early days of RBAIs \cite{r291}, researchers and instructors have raised concerns about whether students might not make a serious attempt at answering the questions on a conceptual-inventory RBAI, such as the FCI, CSEM, or BEMA correctly \cite{r301,r311,r211,r321,r331}. In order for instructors and researchers to appropriately evaluate RBAI data, it is useful to know what proportion of students in a population are taking that RBAI seriously. We define serious students as those who chose answers with consideration, including educated and/or thoughtful guesses, throughout their entire assessment. 

Stewart et al. \cite{r341,r351} study the effect of guessing on both the FCI and CSEM tests. They show that gains are invariant to linear transformation and therefore unaffected by guessing and as such their linear models can correct the test results to account for guessing. Yasuda et al. show that while question 5 scores on the FCI are marginally affected by erroneous reasoning, questions 6, 7, and 16 are more prone to guessing. These questions return a high percentage of false positives as students seem to reach the right answer while using erroneous conceptual reasoning \cite{r361,r371}.

Wang et al. implement item response theory \cite{r381} to build a 3-parameter item response model and use it to analyze student performance on FCI surveys \cite{r391}. They show that a student's proficiency is in linear correlation to a student's raw FCI score. They find that low proficiency students have less than a 5\% chance of guessing the correct answer on questions 23 and 26, and 34\% chance of guess correctly on question 16. They predicts that questions 1 and 6 are the easiest, whereas questions 25 and 26 are the most difficult. As anticipated, each of the 30 questions in the FCI has a different guessing chance and difficulty level, which comes in support of our present work hypothesizing that when students take the survey seriously, there is a better chances that they will select the correct answer for those questions \cite{r391}. 

Hake et al. considered that motivational factors can persuade students to take the RBAIs seriously. Without much evidence at the time, he made the remark that surveyed “students did take the [FCI] pre-test seriously” \cite{r281}. Later, Henderson shows that about 2.8\% of surveyed students may not take an RBAI seriously \cite{r291}.  Henderson was concerned about whether students take the FCI seriously when it is not graded. To identify those students, answer patterns were examined for lack of seriousness from five different angles. By comparison, Pollock et al. ran a longitudinal study of students' conceptual understanding using the BEMA survey, and requested that students report how hard they tried. Three levels were identified: take it very seriously, take it seriously, and did not take it seriously. This study shows that over 50\% of students took the RBAI very seriously, and only 3\% indicated that they did not take it seriously \cite{r401}. 

We have developed a set of seriousness tests and applied them to the FCI, CSEM, and BEMA. It was our goal to develop seriousness tests that could give instructors and researchers an estimate for the proportion of students who did not take an RBAI seriously. Notably, it was not our goal to develop seriousness tests that could identify individual students, and we recommend that the seriousness tests described in this paper not be used in that manner. In subsequent sections we will describe how these seriousness tests were developed as well as those tests' effectiveness in accurately categorizing students as either taking an RBAI seriously or not.

\section{Data sources for the FCI, CSEM, and BEMA}

Data for this paper was obtained from PhysPort's collection of student data. After administering an RBAI, instructors can use the PhysPort Data Explorer to analyze the data from their students. Once the instructors have uploaded their students' responses, the data is stored in a database in PhysPort. We were able to use the data from this database to run our seriousness tests on both the pre- and post-test data for the FCI, CSEM, and BEMA. The database is larger than any data set that has been tested previously, with 64,076 assessment results for the FCI, 15,032 assessment results for the CSEM, and 8,708 assessment results for the BEMA. Table 1 presents the average and the standard deviation for each RBAI.

\begin{table}[]
\begin{tabular}{c|c|c|c|c|}
\cline{2-5}
\multicolumn{1}{l|}{}                            & RBAI & Pre  & Post & Overall \\ \hline
\multicolumn{1}{|c|}{\multirow{3}{*}{Mean (\%)}} & FCI  & 40.5 & 57.3 & 47.8    \\ \cline{2-5} 
\multicolumn{1}{|c|}{}                           & CSEM & 27.8 & 43.8 & 38.9    \\ \cline{2-5} 
\multicolumn{1}{|c|}{}                           & BEMA & 22.8 & 46.1 & 38.7    \\ \hline
\multicolumn{1}{|c|}{\multirow{3}{*}{SD (\%)}}   & FCI  & 20.5 & 22.1 & 22.8    \\ \cline{2-5} 
\multicolumn{1}{|c|}{}                           & CSEM & 12.9 & 18.6 & 18.6    \\ \cline{2-5} 
\multicolumn{1}{|c|}{}                           & BEMA & 10.1 & 18.1 & 19.4    \\ \hline
\end{tabular}
\caption{Means and Standard Deviations for each RBAI}
\label{tab:my-table}
\end{table}

Along with the RBAI results from PhysPort, we created 20,000 simulated RBAI results each for the FCI, CSEM, and BEMA. Our simulated students guessed randomly on all questions. We generated this simulated data in order to model the responses we might expect from non-serious students. Because we could be certain that each simulated individual in the random data set was a random guesser, the seriousness tests needed to flag a significant fraction of this population in order to be considered successful. We did not expect our seriousness tests to identify every member of the simulated population as non-serious, however, because a seriousness test that achieves this would likely lead to misidentifying serious students as non-serious.  It should also be noted that real students are almost never able to behave in a truly random manner on an RBAI, even when they are being non-serious. Their results might show tendencies toward certain answer choices, patterns on the answer sheet, or other trends. This means that students might exist who do not take an RBAI seriously, who also are not well-represented in the simulated population. 

\section{The seriousness tests}
We developed three seriousness tests that can be applied to FCI, CSEM, and BEMA responses in order to estimate the percent of students in a sample who did not take that RBAI seriously: the Pattern Recognition Test (PRT), the Uncommon Answers Test (UAT), and the Easy Questions Test (EQT). These seriousness tests are not designed, however, to identify individual students who did not take an RBAI seriously. In developing these tests, we made the assumption, based on the previous work from Henderson as well as from Pollock et al., that the majority of students take RBAI seriously. As such, we expect the portion of the real population that a successful seriousness test identifies to be small.

\subsubsection{The pattern recognition test}
The Pattern Recognition Test (PRT) is based on the premise that students who do not take an RBAI seriously might choose instead to leave certain patterns throughout their answers. Since computers are not good at picking up on these patterns, we came up with patterns based on what we thought would be likely to find from non-serious test takers. The patterns that we searched for in the RBAIs were:
\begin{itemize}
    \item more than 50\% zeros or blank answers
    \item more than 50\% one letter
    \item 8 of the same letter in a row
    \item 3 instances of ABCD
    \item 2 instances of ABCDE
    \item 1 instance of ABCDEDCBA
\end{itemize}
When these patterns are present in a response, it is likely that the test taker was not taking the RBAI seriously for a significant portion of the test. The correct answers on none of the RBAI evaluated follow any of these patterns. It should also be noted that non-serious students might sometimes produce response patterns outside of those listed above. We limited the patterns that we sought for, however, to avoid misidentifying serious students as non-serious. 

\subsubsection{The uncommon answers test}
The Uncommon Answers Test (UAT) is based on the idea that students who do not take an RBAI seriously sometimes choose answers that were uncommonly chosen by the larger student population. There are nine questions on each of the RBAIs where two or three answer choices were preferred by most of the population. The common answers were most often the correct answer plus one or more of the incorrect answers. Evidently, these preferred choices are attractive to people who were reading carefully through all questions and were being thoughtful in their responses.

If a student chose an unpopular answer on several of these questions, it is likely that they were guessing rather than applying reasoning throughout the assessment. We identified uncommon answer choices based on how few students have picked those answers in the existing PhysPort data. Table 2 summarizes the questions and the less frequently chosen answers. We identified 9 questions with uncommon answers for each RBAI. For the FCI, fewer than 7\% of the population chose one of the uncommon answers for each identified question. For the CSEM, fewer than 10\% of students chose one of the uncommon answers for each identified question. For the BEMA, fewer than 6\% of students chose one of the uncommon answers for each identified question. We counted survey takers who choose at least 4 uncommon answers for the FCI or CSEM or at least 3 uncommon answers for the BEMA as possibly non-serious. 

\begin{table}
\begin{tabular}{|l|c|c|c|c|c|c|c|c|c|}
\hline
RBAI & \multicolumn{9}{c|}{FCI} \\ \hline
Question & 4 & 6 & 12 & 15 & 16 & 22 & 24 & 27 & 29 \\ \hline
Answers & b,c,d & c,d,e & a,d,e & d,e & d,e & c,e & b,d & d,e & c,e \\ \hline
RBAI & \multicolumn{9}{c|}{CSEM} \\ \hline
Question & 1 & 3 & 4 & 7 & 8 & 12 & 13 & 18 & 24 \\ \hline
Answers & a,e & a,e & a,e & d,e & a,e & c,e & c,d & a,b & a,e \\ \hline
RBAI & \multicolumn{9}{c|}{BEMA} \\ \hline
Question & 1 & 2 & 3 & 4 & 5 & 10 & 14 & 21 & 25 \\ \hline
Answers & c,d,g & c,d,g & e,h,i & b,d,j & d,j & h,i & d,e,f & d,j & a,h \\ \hline
\end{tabular}
\caption{Nine questions on each RBAI where a small percentage of the population chose uncommon answers.}
\label{my-label}
\end{table}

\subsubsection{The easy questions test}

The Easy Questions Test (EQT) was based on the idea that students who take a concept-inventory RBAI seriously will get most of the easier questions correct. A student making an effort on such an RBAI might still have one or two of even these questions incorrect, but they are unlikely to be incorrect for all the easy questions. It stands to reason that an answer set in which all the responses to the easy questions are incorrect is more likely to come from a student who did not take that assessment seriously.

We looked at the existing PhysPort data to determine which questions were easiest for students (Figure 1). For each RBAI, we chose the top four questions which had the highest scores, and calculated the percent of students who got a certain number of those questions correct. The students who answered all four easy questions incorrectly were considered as not having taken the assessment seriously. We note, however, that even a random guesser is likely to choose at least one correct answer in any set of five-choice questions. Overall, this means that the EQT will undercount the number of non-serious test-takers. 

\begin{figure}
  \includegraphics[width=0.95 \columnwidth]{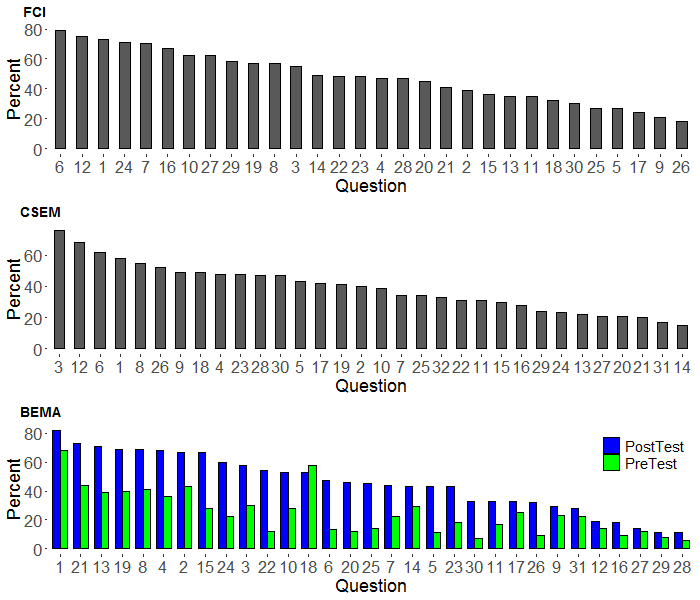}\\
  \caption{Score distribution for each question of each RBAI. The pre- and post-tests scores are combined for the FCI and CSEM.} 
  \label{fig:eo}
\end{figure}

Figure 2 shows the percent of students who answered all of the easy test questions incorrectly for an increasing number of easy questions. We can see that when we choose four or more easy questions, the percent of students who get all of the questions wrong stays relatively constant. For this reason, we chose the four easiest questions from each RBAI based on the proportion of correct responses to that question. The questions chosen for the EQT for each RBAI are shown in Table 3. For the FCI, an easy question has a score greater than 71\%. For the CSEM, an easy question is one with a percent of correct responses above 58\%. As shown in Figure 1 for the BEMA, there is a large discrepancy on the pre- and post-test as to which questions are easy questions. The easy questions for the pre- and post-tests of the BEMA had scores greater than 43\% and 69\%, respectively.

\begin{figure}
  \includegraphics[width=0.95 \columnwidth]{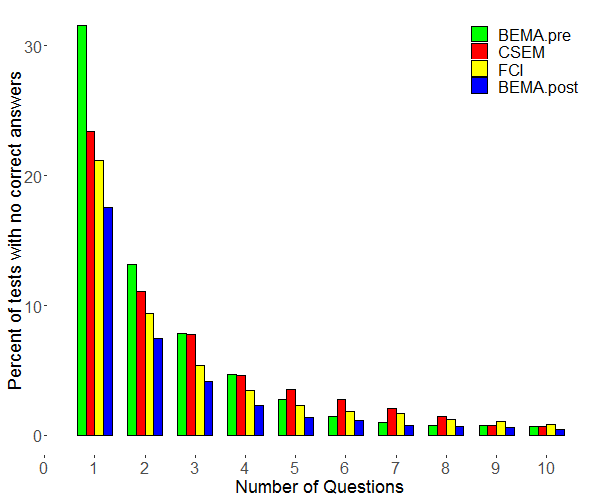}\\
  \caption{Justification for number of Easy Test questions: Percentage of real test takers who answered none of the questions correctly for an increasing number of easy questions as measured by the easy test. The order of the easy test questions was determined by the data in Figure 1.} 
  \label{fig:eo}
\end{figure}

\begin{table}
\begin{tabular}{|l|c|c|c|c|c|}
\hline
RBAI              & \multicolumn{4}{c|}{FCI}  \\ \hline
Question          & 1  & 6  & 12   & 24  \\ \hline
RBAI              & \multicolumn{4}{c|}{CSEM} \\ \hline
Question          & 1  & 3  & 6   & 12     \\ \hline
RBAI              & \multicolumn{4}{c|}{BEMA} \\ \hline
Pre-test Question  & 1  & 2  & 18   & 21  \\ \hline
Post-test Question & 1  & 13  & 19  & 21  \\ \hline
\end{tabular}
\caption{Four questions on each RBAI with the highest percentage of students choosing the right answer.}
\label{my-label}
\end{table}

\section{\label{sec:level1}Analysis and Discussion}

Results of the PRT, UAT, and EQT are shown in Table 4 and in Figure 3. In Figure 3, each of the segments includes the percent of test takers caught by that test, excluding what is shown in the intersecting segments. As an example, the percent of the actual population from the CSEM caught by the PRT is 1.3\%, as shown in Table 4. This comes from combining 0.92\% with each of the segments within the entire PRT circle. The percent of the actual population found to be non-serious by each of these tests is very small, ranging from less than 1\% up to a few percent.

\begin{table}[]
\begin{tabular}{|l|c|c|c|}
\hline
Test & PRT  & UAT & EQT \\ \hline
FCI  & 0.63 & 3.5 & 2.3 \\ \hline
CSEM & 1.3  & 4.6 & 2.5 \\ \hline
BEMA & 1.2  & 3.1 & 3.2 \\ \hline
\end{tabular}
\caption{Percentage of each population detected as non-serious for each RBAI by each of the 3 seriousness tests}
\label{tab:my-table}
\end{table}

\begin{figure*}
  \includegraphics[width=1.75 \columnwidth]{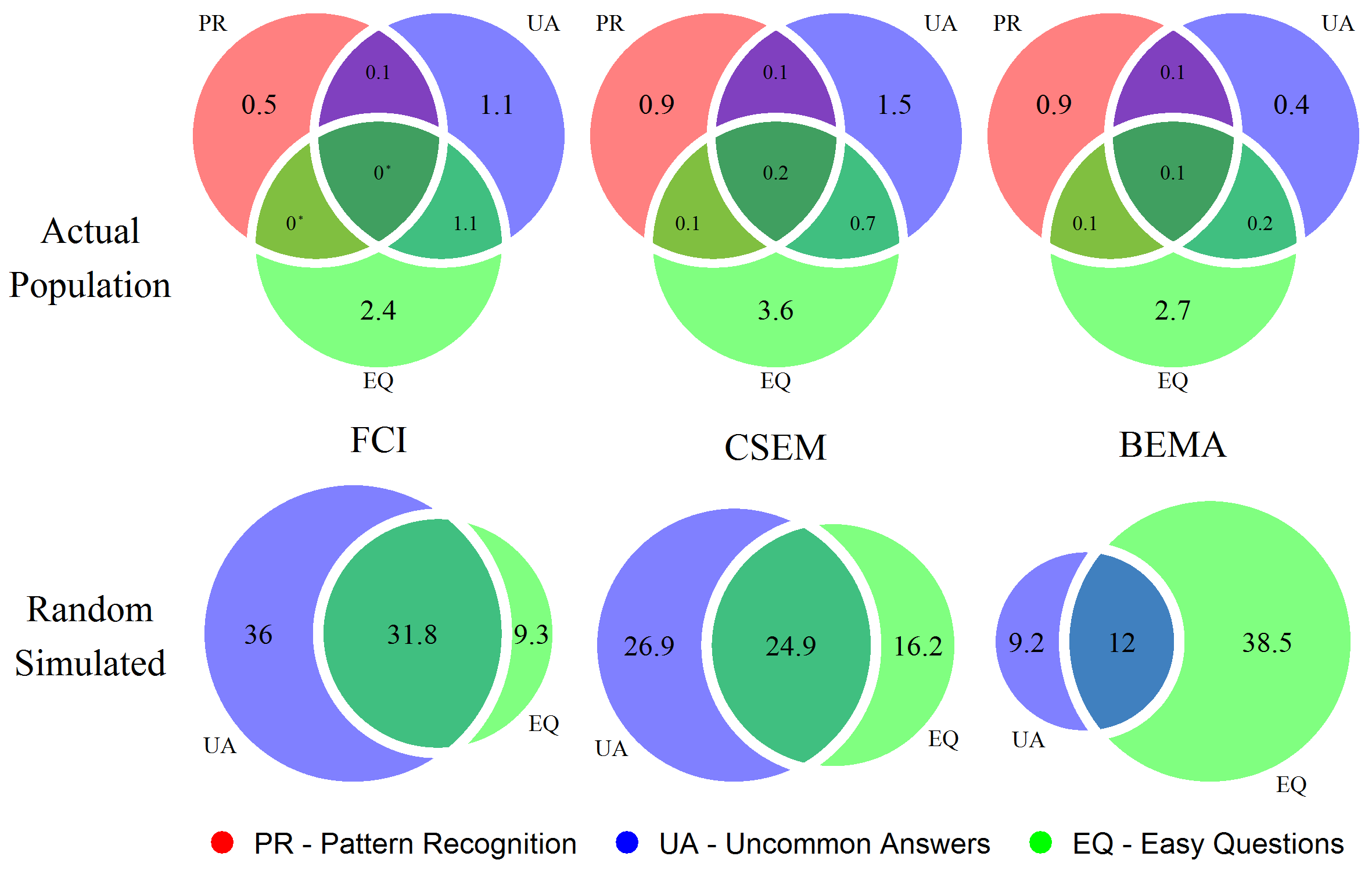}\\
  \caption{Venn diagrams showing the percent of non-serious responses caught by each of the 3 tests for each of the data sets for each RBAI. The percent in each segment is based on the symmetric difference, where each segment includes what is not in the intersection. Note that the PRT results are excluded from the random simulation because almost none of the simulated RBAI results exhibited the patterns sought for by that test. Values with an asterisk round to zero and represent fewer than 32 (FCI), 8 (CSEM), and 5 (BEMA) students.}
  \label{fig:eo2}
\end{figure*}

Applying the PRT to the actual population data identifies between 0.6\% and 1.3\% as non-serious for each of the different RBAIs. The PRT identified nearly zero non-serious survey-takers in the simulated population, however. This is unsurprising because the patterns sought for are non-random. Pattern recognition was thus excluded from Figure 3 for the random simulated results.

For comparison, the UAT and EQT found a high proportion of non-serious responses in the random simulated data. 68\% and 52\% of the respective simulated populations for the FCI and CSEM had four or more Uncommon Answers in their responses, and 21\% of the BEMA random population had three or more Uncommon Answers (blue circle). 41\%, 41\%, and 51\% of the respective simulated populations for the FCI, CSEM, and BEMA were identified by the EQT as non-serious (green circle). In the actual populations, on the other hand, the UAT identified slightly more than 2\% of students as non-serious for each RBAI, and the EQT identified between 3\% and 4.6\% of students as non-serious for each RBAI. 

\begin{figure}
  \includegraphics[width=0.75 \columnwidth]{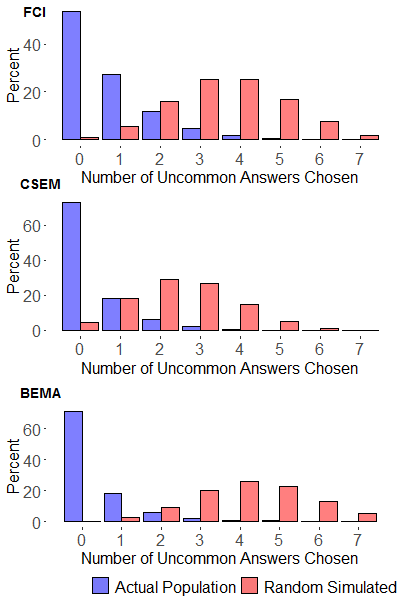}\\
  \caption{Distribution of the percentage of assessments that selected a number of uncommon answers for both the real and simulated populations for each RBAI. There were 9 questions for each RBAI where uncommon answers are rare. Non-serious test takers were those who chose at least 4 uncommon answers for the FCI and CSEM and at least 3 uncommon answers for the BEMA.} 
  \label{fig:eo}
\end{figure}

\begin{figure}
  \includegraphics[width=0.75 \columnwidth]{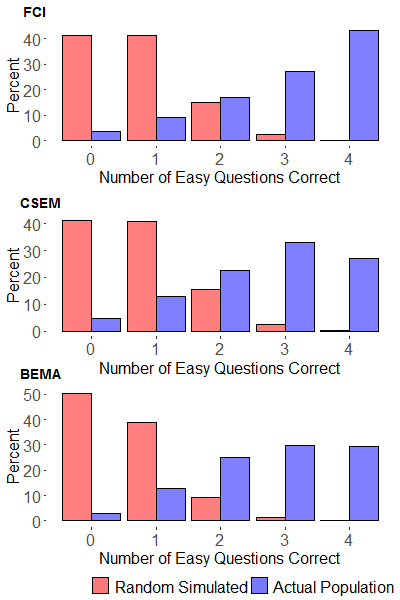}\\
  \caption{Distribution of the percentage of assessments that correctly answered a number of easy questions comparing both the actual and simulated populations for each RBAI. There were 4 questions for each RBAI. Non-serious test takers were those who answered all 4 questions incorrectly.} 
  \label{fig:eo}
\end{figure}

Comparing the uncommon answers chosen for each data set and each RBAI in Figure 4, we see that the actual population chose fewer uncommon answers than the random simulation. Fewer than 50\% of the real students selected any uncommon answers. Conversely, there was an average of three or four uncommon answers for the random population. Therefore, testing for the number of uncommon answers helped us differentiate serious students from random guessers. For the easy questions test, we saw in Figure 5 that the simulated population chose fewer correct easy questions while most of the actual population were able to correctly answer at least 2 easy questions.

\subsubsection*{Combining the results to determine the overall percent of non-serious students}

The percent of the actual population identified by each of the three seriousness tests as non-serious is small, ranging from less than 1\% up to a few percent. The center segments of Figure 3 show that very few real test takers were caught by all three tests for any of the RBAIs, with the FCI catching only 0.016\% with all 3 tests.

As a final comparison, we looked at the scores of any test taker in the actual population who was identified as non-serious by the PRT as well as both of the other two seriousness tests. When we look at the tests caught by the PRT and tests caught by both the EQT and the UAT, the percent of test takers not taking the assessment seriously was 1.5\% to 2.2\% of the population. The number of test takers in the actual population caught by the non-serious tests was 1,120 out of 63,896 assessments (1.8\%) for the FCI, 329 out of 14,876 assessments (2.2\%) for the CSEM, and 133 out of 8,642 assessments (1.5\%) for the BEMA. These values could be determined using Figure 3 by combining the percentages within the entire red (PR) circle with the overlapping segment between UA and EQ. These results were very similar, suggesting that our seriousness tests accurately determine the percent of students who did not take an RBAI seriously. Figure 6 shows graphs of the non-serious scores, identified in the combined manner described vs. all of the scores in the actual population. From this graph, we can see that the non-serious assessment scores are much lower than those from the actual population. This is further evidence that the scores identified by a combination of the PRT and both the UAT and the EQT were most likely truly not serious.

\begin{figure}
  \includegraphics[width=0.95 \columnwidth]{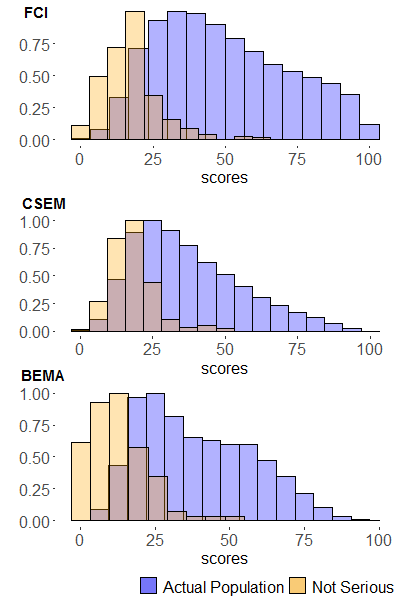}\\
  \caption{All vs. Not-Serious Scores: Distribution of scores for each RBAI comparing all of the scores in the actual population to the responses caught by the tests and considered a non-serious attempt by the actual population.} 
  \label{fig:eo}
\end{figure}

\section{\label{sec:level1}Conclusion}

Our results are in contrast to work mentioned in the introduction by Henderson, who found that about 2.8\% of students did not take the FCI seriously \cite{r301}, and the results from by Pollock et al. who found that 3\% indicated that they did not take the BEMA seriously \cite{r401}. We find that fewer students were caught by our seriousness tests, and we conclude that the overall percentage of students who did not take the CIs seriously is only about 1.5\% - 1.6\%. Regardless, our results are in line with previous work that shows that the incidence of non-seriousness in RBAI results is very low. 

In addition, our seriousness tests might undercount incidents of non-seriousness. We made deliberate choices to avoid misidentifying serious test-takers as non-serious, and those choices could have resulted in misidentifying some non-serious test-takers as serious. It is still likely, however, that the methods described will sometimes falsely identify a serious student as non-serious. Because of this, we do not recommend using any of these seriousness tests to identify individual students as serious or non-serious.  

We suggest that the three seriousness tests developed here could be used together as described to give reasonable estimates of percents of non-seriousness for FCI, CSEM, and BEMA datasets, and that results similar to those just described indicate a low incidence of non-seriousness in a dataset. In addition, these seriousness tests might be applied to other concept-inventory RBAIs, although some details would need to be worked out for the UAT and EQT for each RBAI.

\end{document}